\pdfoutput=1
\documentclass{article}

\PassOptionsToPackage{numbers,compress}{natbib}
\usepackage[final,sglblindworkshop]{neurips_2026}

\usepackage[utf8]{inputenc}
\usepackage[T1]{fontenc}
\usepackage{amsmath,amsfonts,url}
\usepackage{hyperref}
\hypersetup{hypertexnames=false,
  pdftitle={Harmonic Circuit: Chord Progression Lies in a Sparse Principal Information Flow in Symbolic Music Models},
  pdfauthor={Ruiji Yu, Gus Xia}}
\usepackage{graphicx}
\usepackage{xcolor}
\usepackage{float}
\usepackage{placeins}
\usepackage{booktabs}
\usepackage{multirow}
\usepackage{tabularx}
\usepackage{etoolbox}
\usepackage{xurl}
\usepackage{microtype}
\usepackage{tikz}
\usetikzlibrary{arrows.meta,decorations.pathreplacing,positioning}

\definecolor{abcField}{HTML}{F3C6B5}
\definecolor{abcChord}{HTML}{E5F1E1}
\definecolor{abcNote}{HTML}{E7EEF6}
\definecolor{abcAccent}{HTML}{EF7D4F}
\definecolor{abcNoteAccent}{HTML}{3B82C4}
\definecolor{abcBrace}{HTML}{36A84A}
\definecolor{abcText}{HTML}{30343F}
\definecolor{abcTarget}{HTML}{F6D55C}

\graphicspath{{./}}
\newcolumntype{Y}{>{\raggedright\arraybackslash}X}

\title{Harmonic Circuit: Chord Progression Lies in a Sparse Principal Information Flow in Symbolic Music Models}
\workshoptitle{Interpretability for Discovery}

\author{%
  Ruiji Yu \qquad Gus Xia \\
  Mohamed bin Zayed University of Artificial Intelligence \\
  \texttt{\{ruiji.yu, gus.xia\}@mbzuai.ac.ae} \\
}

\begin{document}

\maketitle

\begin{abstract}
Mechanistic Interpretability (MI) seeks to decode neural network computations into human-understandable logic. A core methodology in MI is circuit discovery: identifying the information flow in sub-networks that drive specific model behaviors. While circuit discovery has matured within Natural Language Processing (NLP), the interpretable circuits in music models remain largely unexplored. In this work, we identify a music-functional circuit within two symbolic music generative models: ChatMusician and MuPT. The music function we consider is the prediction of the terminal tonic root in a classical harmonic progression pattern $I\ldots V\rightarrow I$. We use the Auto-Emergence algorithm to extract a principal sparse sub-network as the identified circuit. Validation via causal intervention confirms that this identified circuit is a primary driver of harmonic progression behaviors in the symbolic music models. To our knowledge, this represents the first systematic circuit-discovery study conducted within the music domain. Our demo page is available at: \url{https://ukiyo23rui.github.io/auto-emergence-demo/}.
\end{abstract}

\section{Introduction}\label{sec:introduction}

AI models have become increasingly capable in both music understanding and generation~\cite{Chatmusician,Notagen,MusicGen,Musicbert,Wholesong}, driven by scaling laws and the rapid development of training data. However, there is limited understanding of how the internal components in these models coordinate to interpret and process complex musical patterns, such as harmonic progressions.

Mechanistic interpretability techniques aim at translating black-box processes inside neural networks, e.g., learned features and computations, into algorithms and concepts that are interpretable to humans~\cite{Mechanistic_Survey,Probing,SAE,Circuit_IOI,Information_Flow_Routes,same}, advancing the development of AI systems that are trustworthy and beneficial to people. In mechanistic interpretability, a \textbf{circuit} is defined as a human-interpretable sub-network of the model that is responsible for some specific behavior~\cite{Circuit_IOI}.

\begin{figure}[!t]
  \centering
  \includegraphics[width=0.95\linewidth]{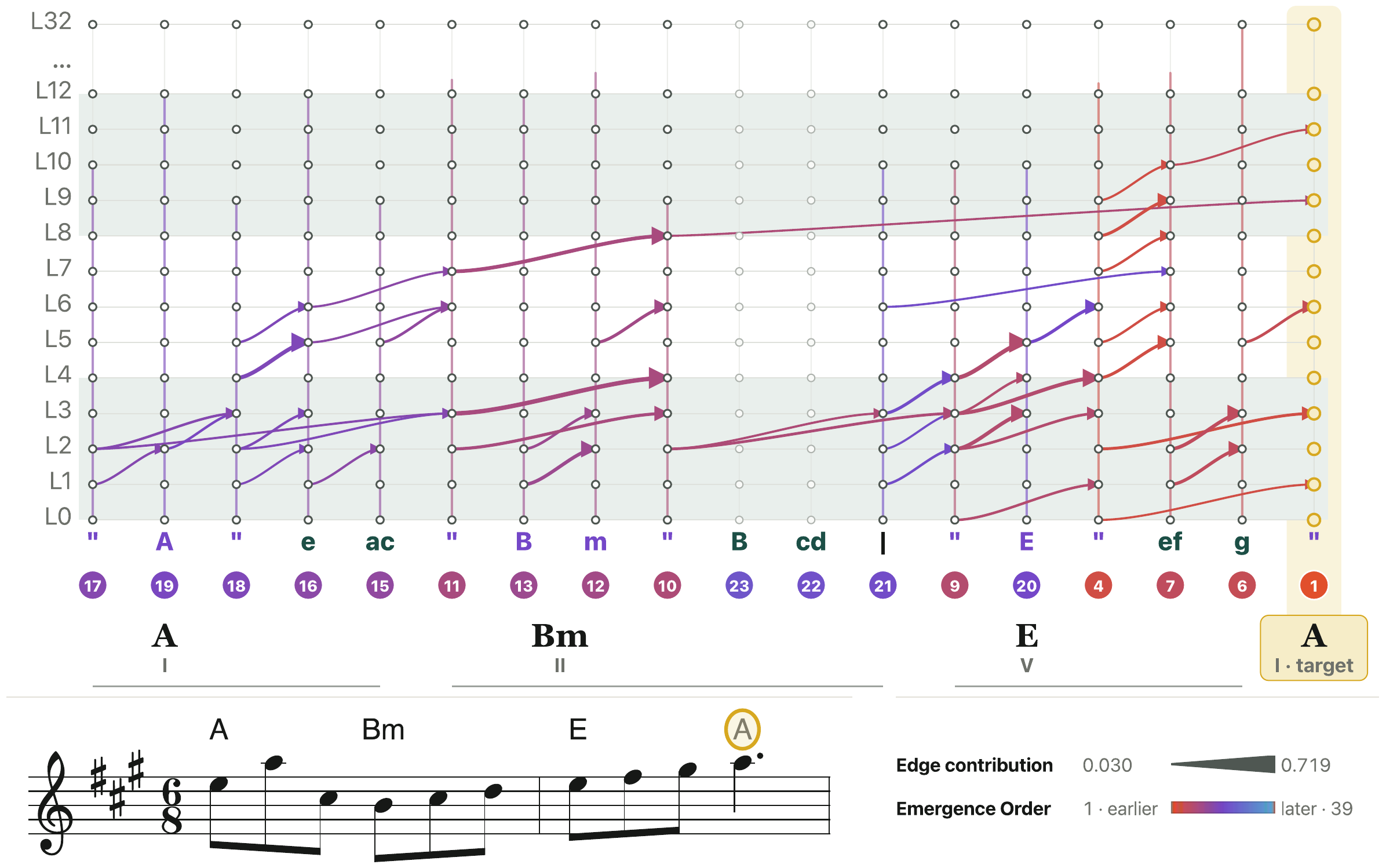}
  \caption{Principal information flow circuit in a harmonic-progression example. The horizontal axis shows the music tokens given to the model, and the vertical axis shows the Transformer layers. The yellow highlight marks the prediction goal of tonal resolution. Emergence Order ranks the music tokens by when they first enter the recursively extracted principal information flow circuit.}
  \label{fig:intro_information_flow}
\end{figure}

In the music domain, several works have made meaningful attempts in mechanistic interpretability. However, almost all of them remain confined to the representational level of explanation, such as probing~\cite{Music_Activation,Music_Probing} or SAEs~\cite{Music_SAE}. To our knowledge, no prior work has attempted to discover circuits, which are important for understanding how models organize their internal components to dynamically interact with the naturally structured music data.

An intuitive question, then, is what kind of musical behavior naturally lends itself to a circuit, implying the emergence of a structured sub-network within the model. We argue that the harmonic chord progression is a strong candidate because it is governed by an explicit structure, which may lead the model to form internally differentiated components for handling distinct harmonic roles.

A useful analogy is the Indirect Object Identification (IOI)\footnote{In IOI, given two names in a sentence, the model predicts the name that is not the subject of the last clause. For example, in ``When Mary and John went to the store, John gave a drink to'', the correct completion is ``Mary''.} task~\cite{Circuit_IOI}, where a language model needs to distinguish the subject and the object in the sentence. Harmonic progression provides a similar syntax-guided form of dependency in music. In the familiar functional pattern $I\ldots V\rightarrow I$ (this notation is explained in Section~\ref{subsec:music_primer}), the first tonic establishes the center, the dominant builds tension, and the final tonic resolves it. Intuitively, to predict the terminal resolution correctly, the model needs to preserve and route harmonically relevant information across positions. Therefore, this harmonic-resolution behavior is a suitable target for circuit discovery.

We conduct circuit discovery using music pieces with an $I\ldots V\rightarrow I$ chord progression extracted from Nottingham~\cite{Nottingham} and use a subset of the Bach Chorales Melody--Harmony Corpus (BCMH), selected by the same chord criterion, to test polyphonic generalization~\cite{BCMH_Dataset,Sears2023Order}. We first use Information Flow Routes to construct layer-wise token contribution matrices~\cite{Information_Flow_Routes} and recursively extract the principal information flow circuit with Auto-Emergence~\cite{Information_Flow_Trust}, shown in Figure~\ref{fig:intro_information_flow}. Finally, we validate the circuit with intervention.

Our study takes a first step toward bringing circuit analysis into the music domain. While we currently focus on a single, highly specified task centered on one harmonic behavior, we hope that this line of analysis will not only deepen our understanding of music models but also enable us to learn from the musical knowledge encoded in music models, such as composition motifs, in the future.

\section{Background: harmony and music notation}\label{subsec:music_primer}

\paragraph{Keys, chords, and harmonic resolution.}
Tonal music is organized around a \emph{key}, which fixes a central pitch and the scale built on it; the position of a note within this scale is its \emph{scale degree}. A \emph{chord} is a set of notes played simultaneously, and the note on which it is built is its \emph{root}. Chords are named by the scale degree of their root, written as Roman numerals (lower case for minor chords). In A major, for example, $I$ is the chord on A, the \emph{tonic}, and $V$ is the chord on the fifth degree E, the \emph{dominant}. In this paper, $I$ and $V$ refer only to the scale degree of the root, regardless of chord quality, so that one notation covers major and minor keys. The tonic is heard as stable and the dominant as tense, and the motion $V\rightarrow I$, called \emph{resolution}, is the canonical way to close a phrase~\cite{OMT_CadenceTypes,Caplin2024Cadence,CheongDing2018Sposobin}. We study the chord progression $I\ldots V\rightarrow I$: a passage that starts on a tonic-root chord, passes through any number of intermediate chords, and ends with a dominant-root chord immediately followed by a tonic-root chord.

\paragraph{Monophonic and polyphonic music.}
A \emph{monophonic} piece has a single melodic line, possibly annotated with chord symbols, as in a folk-tune lead sheet. A \emph{polyphonic} piece has several lines sounding at once, such as the four voices of a chorale, from high to low soprano, alto, tenor, and bass (SATB).

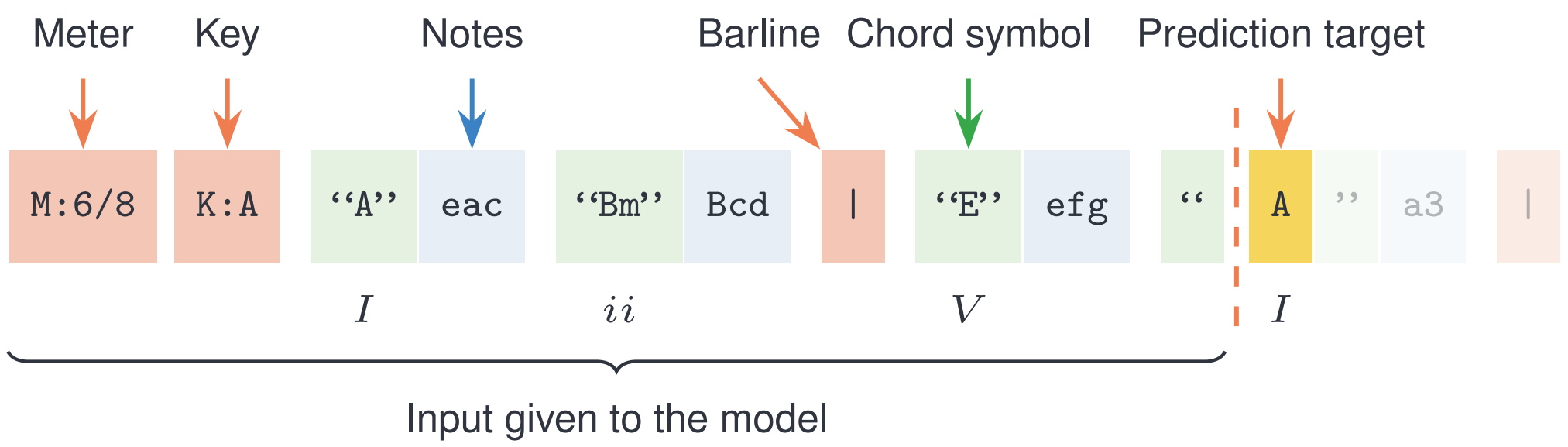
\begin{figure}[t]
  \centering
  \resizebox{0.95\linewidth}{!}{%
  \begin{tikzpicture}[
    token/.style={font=\ttfamily\tiny, text=abcText, minimum height=0.6cm,
      text height=4pt, text depth=1.25pt,
      inner xsep=3.2pt, inner ysep=1.5pt, anchor=west},
    withheld/.style={opacity=0.35, text opacity=0.35},
    callout label/.style={font=\sffamily\tiny, text=abcText,
      text height=4pt, text depth=1.25pt},
    callout/.style={-{Stealth[length=2.1mm,width=1.5mm]}, draw=abcAccent,
      line width=0.7pt}
  ]
    \node[token, fill=abcField] (meter) at (0,0) {M:6/8};
    \node[token, fill=abcField, right=0.8mm of meter] (key) {K:A};
    \node[token, fill=abcChord, right=1.5mm of key] (chordA) {``A''};
    \node[token, fill=abcNote, right=0mm of chordA] (notesA) {eac};
    \node[token, fill=abcChord, right=1.5mm of notesA] (chordBm) {``Bm''};
    \node[token, fill=abcNote, right=0mm of chordBm] (notesBm) {Bcd};
    \node[token, fill=abcField, right=1.5mm of notesBm] (barOne) {|};
    \node[token, fill=abcChord, right=1.5mm of barOne] (chordE) {``E''};
    \node[token, fill=abcNote, right=0mm of chordE] (notesE) {efg};
    \node[token, fill=abcChord, right=1.5mm of notesE] (finalOpen) {``};
    \node[token, fill=abcTarget, right=1.2mm of finalOpen] (finalRoot) {A};
    \node[token, fill=abcChord, withheld, right=0mm of finalRoot] (finalClose) {''};
    \node[token, fill=abcNote, withheld, right=0mm of finalClose] (notesFinalA) {a3};
    \node[token, fill=abcField, withheld, right=1.5mm of notesFinalA] (barTwo) {|};

    \node[callout label, above=0.375cm of meter] (meterLabel) {Meter};
    \node[callout label, above=0.375cm of key] (keyLabel) {Key};
    \node[callout label, above=0.375cm of notesA] (notesLabel) {Notes};
    \node[callout label, above=0.375cm of chordE] (chordLabel) {Chord symbol};
    \node[callout label, above=0.375cm of barOne, xshift=-5mm] (barLabel) {Barline};
    \node[callout label, above=0.375cm of finalRoot] (targetLabel) {Prediction target};
    \draw[callout] (meterLabel.south) -- (meter.north);
    \draw[callout] (keyLabel.south) -- (key.north);
    \draw[callout, draw=abcNoteAccent] (notesLabel.south) -- (notesA.north);
    \draw[callout, draw=abcBrace] (chordLabel.south) -- (chordE.north);
    \draw[callout] (barLabel.south) -- (barOne.north west);
    \draw[callout] (targetLabel.south) -- (finalRoot.north);

    \draw[dashed, draw=abcAccent, line width=0.7pt]
      ([xshift=0.6mm,yshift=2.2mm]finalOpen.north east) --
      ([xshift=0.6mm,yshift=-4.2mm]finalOpen.south east);
    \draw[decorate, decoration={brace, mirror, amplitude=3pt},
      draw=abcText, line width=0.5pt]
      ([yshift=-4.6mm]meter.south west) -- ([yshift=-4.6mm]finalOpen.south east)
      node[midway, below=1.6mm, callout label] {Input given to the model};
    \node[callout label, below=0.4mm of chordA] {$I$};
    \node[callout label, below=0.4mm of chordBm] {$ii$};
    \node[callout label, below=0.4mm of chordE] {$V$};
    \node[callout label, below=0.4mm of finalRoot] {$I$};
  \end{tikzpicture}%
  }
  \caption{An annotated Nottingham excerpt in ABC notation. The key is A major and the chord symbols A--Bm--E--A form the progression $I$--$ii$--$V$--$I$. The model receives everything to the left of the dashed line and must predict the root of the final chord (highlighted); the faded tokens are withheld.}
  \label{fig:nottingham_abc_sample}
\end{figure}

\paragraph{ABC notation.}
Both models in this study read and write music as plain text in ABC notation. A tune starts with header fields such as the meter (\texttt{M:6/8}) and the key (\texttt{K:A}); notes are written as letters, a vertical bar \texttt{|} separates measures, and a chord symbol is written in double quotes before the notes it accompanies, e.g., \texttt{"Bm"} for a B minor chord. Figure~\ref{fig:nottingham_abc_sample} shows an annotated Nottingham example~\cite{Nottingham}. Polyphonic pieces use the same format, with the four voices of each chord serialized one after another (Figure~\ref{fig:bcmh_case_intervention} in Appendix~\ref{app:data_processing_details}).

\section{Methodology}\label{sec:methodology}

We adopt two existing information-flow tools. Information Flow Routes~\cite{Information_Flow_Routes} converts the symbolic music generative model's forward computation into a weighted, layer-wise graph, and Auto-Emergence~\cite{Information_Flow_Trust} extracts the \emph{principal information flow circuit}, which we abbreviate as the \emph{principal circuit}. We then use intervention to validate the effect of the principal circuit.

\subsection{Task and data representation}\label{subsec:data_representation}

\paragraph{Representation.}
We treat the complete quoted span of a chord symbol as a harmonic site and retain notes, barlines, meter, key, and other symbols as musical context. Because tokenization is model-specific, a chord symbol may correspond to one token or multiple tokens; we align the annotated chord root to the tokenizer span that contains its root character, so that chord-symbol sites and root-degree roles can be mapped to model-token positions.

\paragraph{Task.}
The model is given the beginning of a piece in ABC notation, ending just before the root of the final chord of an $I\ldots V\rightarrow I$ progression, and it must predict that root as its next token. In the example of Figure~\ref{fig:nottingham_abc_sample}, the key is A major and the chords so far are A, Bm, and E, i.e., $I$, $ii$, and $V$. The input ends after the opening quotation mark of the next chord symbol, and the correct continuation is the letter \texttt{A}, the root of the tonic chord.

\subsection{Information flow graph modeling}\label{subsec:information_flow_graph_modeling}

\begin{figure}[t]
  \centering
  \includegraphics[width=\linewidth]{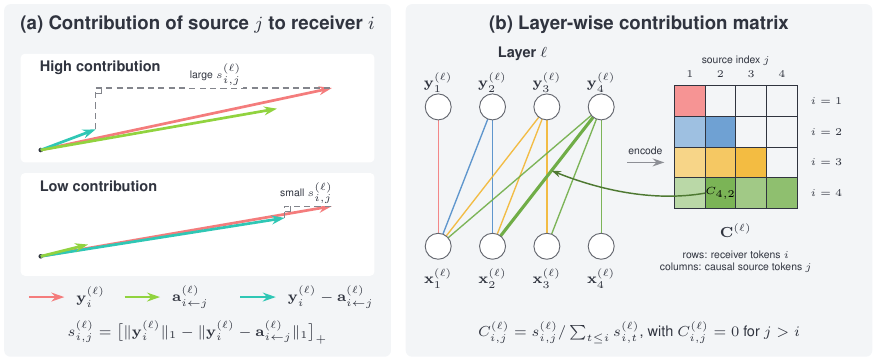}
  \caption{Information flow graph modeling.
  (a) Node-level contribution of source $j$ to receiver $i$.
  (b) Layer-wise contribution matrix $\mathbf{C}^{(\ell)}$.}
  \label{fig:contribution_definition}
\end{figure}

Information Flow Routes~\cite{Information_Flow_Routes} analyzes how Transformer-based models organize the contextual information for next-token prediction. It treats each input representation at every layer as an individual node. At each layer $\ell$, the sum of all attention head outputs from position $j$ to position $i$ is defined as \textbf{an attribution vector} $\mathbf{a}^{(\ell)}_{i\leftarrow j}$, which represents the directed edge $j\to i$. To quantify the strength of each directed edge $j\to i$, we define its raw contribution as the nonnegative decrease in the $L_1$ norm of the receiver state $\mathbf{y}^{(\ell)}_i$ when the corresponding attribution vector is removed, as illustrated in Figure~\ref{fig:contribution_definition}. We then normalize these scores across all causal source positions $j\leq i$, yielding the lower-triangular \textbf{contribution matrix} $\mathbf{C}^{(\ell)}$. Each $C^{(\ell)}_{i,j}$ therefore represents the relative contribution of source $j$ to the state update at receiver $i$; we call the edges with $j<i$ \emph{cross-token edges} and write $(\ell,i,j)$ for such an edge. Repeating this construction across all token positions and layers yields a weighted information-flow graph for the current prediction.

\subsection{Principal information flow circuit extraction}\label{subsec:recursive_principal_subgraph_extraction}

We apply Auto-Emergence~\cite{Information_Flow_Trust} to this weighted graph to extract the principal circuit. This algorithm initializes the principal circuit with the residual representations of the last input position at all layers and adds all of their incoming cross-token edges to a selection pool. It then chooses the strongest available edge $C^{(k)}_{i,j}$ in the pool and adds it to the principal circuit. Every time an edge is selected, the algorithm retains the residual representation of the source token $j$ below layer $k$ and adds all of its incoming edges from lower layers to the selection pool. The algorithm recursively expands the selection process until all tokens have been added to the principal circuit. The order in which the tokens are added is their \emph{emergence order}. The selected edges are returned as sparse matrices $\{\mathbf{P}^{(\ell)}\}_{\ell=1}^{L}$; Appendix~\ref{app:auto_emergence} gives the details.

In our analysis, the prediction target is the terminal tonic root in each $I\ldots V\rightarrow I$ sequence. The input is truncated before this target token. The selected edges of the principal circuit show which earlier harmonic sites contribute to the terminal root prediction and at which layers they enter the principal circuit. Because the extraction starts from one prediction, every sample $s$ has its own principal circuit; we write $\mathcal{U}_s$ for the set of all cross-token edges of the sample and $\mathcal{E}_s\subseteq\mathcal{U}_s$ for the edges selected into its principal circuit.

\subsection{Causal intervention validation}\label{subsec:causal_intervention_validation}

We use attention-edge intervention~\cite{Circuit_IOI} to test whether the extracted principal circuit is necessary for predicting the terminal tonic. Let $A^{(\ell,h)}_{i,j}$ denote the post-softmax attention probability from source position $j$ to receiver position $i$ in head $h$ of layer $\ell$, where $j<i$. For each selected edge $(\ell,i,j)$, we set
\begin{equation}
A^{(\ell,h)}_{i,j}\leftarrow 0\qquad\text{for all heads }h,
\end{equation}
without renormalizing the remaining attention probabilities. In our intervention, we simultaneously ablate all selected cross-token edges of the principal circuit and then measure the effect on the model's prediction of the final tonic. As a \emph{random control}, we ablate the same number of edges drawn uniformly at random from $\mathcal{U}_s$.

\section{Experiments}\label{sec:experiments}

\subsection{Datasets}\label{subsec:datasets}
Our main experiments use 963 $I\ldots V\rightarrow I$ chord pattern sequences (882 in major keys and 81 in minor keys) extracted from the Nottingham ABC corpus~\cite{Nottingham}, a collection of folk tunes with chord symbols. We also construct 98 four-part samples with soprano, alto, tenor, and bass (SATB) voices from BCMH~\cite{BCMH_Dataset}, with the same chord pattern selection criterion, to test the circuit's generalization to polyphonic music. Appendix~\ref{app:data_processing_details} gives the processing details.

\subsection{Models}\label{subsec:models}
We evaluate ChatMusician-7B~\cite{Chatmusician}, a LLaMA-2-based language model adapted to symbolic music through continual pretraining and fine-tuning on ABC-based music and text data. We also evaluate MuPT-190M~\cite{MuPT}, a symbolic music model pretrained specifically on large-scale monophonic and polyphonic corpora. Both models read and generate music as ABC text. Both models are tested on Nottingham; we use only MuPT for BCMH because its pretraining explicitly includes polyphonic music.

\subsection{Evaluation metrics}\label{subsec:evaluation_metrics}
At the prediction position of the target root, we measure the effect of intervening on the principal circuit. In the full-vocabulary setting, we report the target tonic token's \textbf{logit} and prediction \textbf{rank}, the \textbf{top-1 accuracy} (the percentage of samples for which the target tonic token has the highest probability among all vocabulary tokens), and the \textbf{KL divergence} between the pre- and post-intervention vocabulary distributions. To focus on the harmonic choice, we further restrict the candidates to the twelve pitch classes, merging enharmonic spellings such as C$\sharp$ and D$\flat$, and compute twelve-root top-1 accuracy, target rank, KL divergence, and the \emph{twelve-root margin}, i.e., the score of the target root minus that of the highest-scoring competing root (Appendix~\ref{app:twelve_root_metrics}).

\subsection{Principal information flow circuit intervention results}\label{subsec:principal_information_flow_intervention_results}
For each sample, we apply the intervention of Section~\ref{subsec:causal_intervention_validation} and compare it with the mean of five same-size random controls. In every setting, the principal circuit contains fewer than 6\% of all cross-token edges.

\begin{figure}[t]
  \centering
  \includegraphics[width=\linewidth]{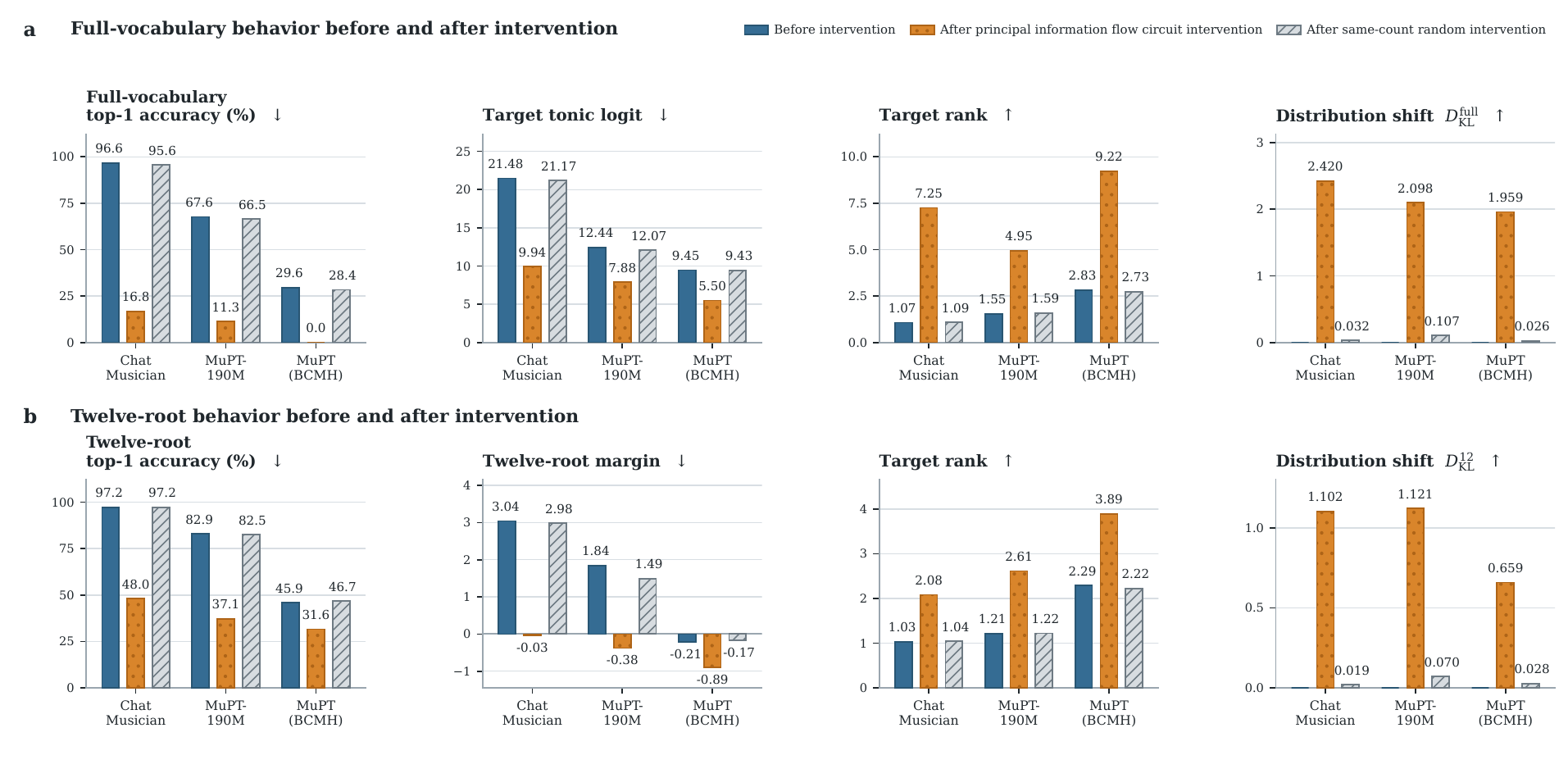}
  \caption{Evaluation before and after principal information flow circuit intervention for
  (a) full-vocabulary and (b) twelve-root prediction with enharmonic pairs merged.
  Metrics are top-1 accuracy ($\downarrow$), target logit or twelve-root margin
  ($\downarrow$), target rank ($\uparrow$), and KL divergence ($\uparrow$); arrows indicate stronger
  intervention effects. Blue, orange, and gray bars denote original, circuit-intervened, and
  same-count random-edge-intervened values, respectively. The circuit intervention consistently
  produces the larger effect.}
  \label{fig:intervention_metric_comparison}
\end{figure}

Figure~\ref{fig:intervention_metric_comparison} supports three conclusions. First, the principal circuit is necessary for the tonic prediction: ablating it removes most of both models' ability to predict the correct tonic, whereas ablating the same number of random edges leaves the prediction almost unchanged. Second, the ablation of the principal circuit pushes the models toward other outputs: the correct tonic falls behind several competing tokens, and the whole output distribution shifts. The same holds when only the twelve possible chord roots are compared, so the effect lies in which chord root the models choose. Third, the circuit remains important in polyphonic music: on BCMH, where MuPT's baseline is considerably lower, ablating the circuit still lowers twelve-root accuracy well below that of the random control.

\paragraph{Consistency and composition of the circuit.}
The \emph{margin} of a sample is how far the correct root letter's logit is ahead of the best competing letter; the \emph{net margin drop} of a set of edges is the difference between the margin after ablating the same number of random edges and the margin after ablating the set, so a positive value means that the set matters more than random edges (Appendix~\ref{subsec:root_letter_margin}). Ablating the circuit is more damaging than the random control in almost every sample for ChatMusician and in about three quarters of the samples for MuPT (Appendix~\ref{sec:circuit_analysis}). Per edge, the effect is strongest inside chord symbols: these edges form only a small part of the circuit, yet at an equal number of ablated edges they matter far more than any other category of edges (Figure~\ref{fig:edge_family_summary}; the five edge categories are defined in Appendix~\ref{subsec:edge_families}). This suggests that both models first assemble a representation of each chord within its own symbol and rely on it for the final prediction.

\begin{figure}[t]
  \centering
  \includegraphics[width=\linewidth]{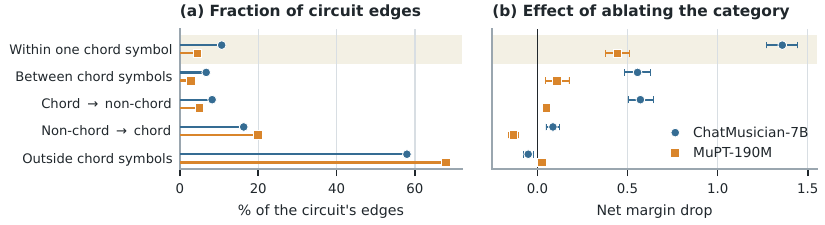}
  \caption{Edge categories of the principal circuit on Nottingham. (a) Fraction of the circuit's cross-token edges that belong to each category. (b) Net margin drop when the same number of edges is ablated from each category, with 95\% bootstrap confidence intervals.}
  \label{fig:edge_family_summary}
\end{figure}

\section{Conclusion and limitations}\label{sec:limitations}
This work provides a first systematic step toward information flow analysis and circuit discovery in the music domain. We extract and validate principal information flow circuits for harmonic chord progression in ChatMusician and MuPT. The study has two main limitations. First, the principal circuit is extracted for each prediction separately, so our findings describe regularities across the set of circuits; we do not identify a single fixed circuit, such as a list of attention heads, that is shared by all samples. Second, the task is relatively narrow and single: we study one harmonic behavior with syntactic structure, the prediction of the terminal tonic root in $I\ldots V\rightarrow I$ progressions, on Western tonal music. In the future, we will extend our analysis to more complex musical behaviors and investigate more detailed functional components of the discovered circuits.

\begin{ack}
We thank Daniel Chin and Liwei Lin, PhD students at Music X Lab, for their suggestions on the writing and polishing of this paper.
\end{ack}

\section*{Responsible use statement}
This work studies the mechanistic interpretability of existing symbolic music models. Its potential benefits include improved transparency and interpretability of music-generation systems. Its potential risks include overgeneralizing the findings beyond the evaluated Western tonal corpora. We mitigate these risks by limiting our claims to the evaluated models, datasets, and harmonic task, by stating the limitations of the study, and by releasing artifacts subject to the licenses of the original models and datasets.

\bibliographystyle{plainnat}
\bibliography{references}

\clearpage
\appendix
\raggedbottom

\section{Related work}\label{sec:related_work}

\subsection{Autoregressive symbolic music generation}\label{subsec:autoregressive_symbolic_music_generation}
Autoregressive symbolic music models represent a score as a sequence of discrete tokens and generate musical events sequentially according to $p_\theta(x_{1:T})=\prod_{t=1}^{T}p_\theta(x_t\mid x_{<t})$. Early Transformer-based architectures, including Music Transformer and Museformer, explored the modeling of long-range musical structure~\cite{MusicTransformer,Museformer}. Recently, ChatMusician and MuPT have adopted large-scale pretraining and language-model architectures. NotaGen further incorporates preference optimization to improve generation quality and controllability~\cite{Chatmusician,MuPT,Notagen}. Autoregressive models all center on next-token prediction, so we can analyze the prediction performance at target-root positions and investigate the causal effect of internal information flow in harmonic prediction.

\subsection{Mechanistic interpretability and circuit discovery}\label{subsec:mechanistic_interpretability_circuit_discovery}
Mechanistic interpretability aims to reverse-engineer computational mechanisms learned by neural networks into human-understandable algorithms and concepts~\cite{Mechanistic_Survey}. Circuit discovery has become a central paradigm in this area, with foundations in the Transformer Circuits framework~\cite{Circuit_mathematical} and concrete successes on benchmark behaviors such as IOI~\cite{Circuit_IOI} and Greater-Than~\cite{Greater}. Early workflows mainly relied on activation/path patching and component-wise causal intervention~\cite{Circuit_IOI,Locating}, which can be costly at scale. More recent automated or attribution-based approaches, including ACDC~\cite{ACDC}, Information Flow Routes~\cite{Information_Flow_Routes}, and Attribution Patching~\cite{Attribution}, have improved the practicality of circuit analysis. Nevertheless, circuit discovery remains concentrated in NLP, while its application to music is still underexplored. Starting from a concrete harmonic chord progression behavior, our work takes a step toward extending circuit discovery to music.

\section{Auto-Emergence extraction details}\label{app:auto_emergence}

Auto-Emergence~\cite{Information_Flow_Trust} takes the lower-triangular contribution matrices $\{\mathbf{C}^{(\ell)}\}_{\ell=1}^{L}$ of an input with $T$ tokens and returns sparse principal matrices $\{\mathbf{P}^{(\ell)}\}_{\ell=1}^{L}$ and an emergence order $\mathbf{E}\in\mathbb{R}^{T}$, in which a smaller nonzero $E_j$ means that token $j$ entered the principal circuit earlier. It starts from the prediction position $T$: it sets $E_T=1$, keeps $P^{(\ell)}_{T,T}=C^{(\ell)}_{T,T}$ for every layer, and initializes the selection pool $\mathcal{S}=\{C^{(\ell)}_{T,j}\mid j<T,\ 1\leq\ell\leq L\}$ with all cross-token edges entering $T$. It then repeats two steps until every token has an emergence rank:
\begin{enumerate}
  \item Select the strongest edge in the pool, $C^{(k)}_{i,j}=\max\mathcal{S}$, set $P^{(k)}_{i,j}=C^{(k)}_{i,j}$, remove it from $\mathcal{S}$, and give source $j$ the next rank if it has none.
  \item Because this edge uses the representation of $j$ at layer $k$, keep $P^{(\ell)}_{j,j}=C^{(\ell)}_{j,j}$ for $\ell<k$ and add the edges that feed it, $\{C^{(\ell)}_{j,m}\mid m<j,\ \ell<k\}$, to $\mathcal{S}$.
\end{enumerate}

\section{Twelve-root evaluation details}\label{app:twelve_root_metrics}

For each sample, we preserve the observed chord suffix (quality, extension, inversion, and closing quote) and replace only its root with each of 17 conventional spellings from \texttt{C}, \texttt{C\#}, \texttt{Db}, \ldots, \texttt{Bb}, and \texttt{B}. Because a candidate may contain one or more model-specific tokens, its score is the teacher-forced sum of conditional log-probabilities from the root onset through the closing quote,
\begin{equation}
s(r)=\sum_{t=t_0}^{t_1}\log p_\theta\!\left(x_t^{(r)}\mid
x_{<t}^{(r)}\right).
\end{equation}
We map the 17 spellings to 12 pitch classes and use the highest-scoring enharmonic spelling as the primary class score, $\widetilde{s}_c=\max_{r:\,\operatorname{pc}(r)=c}s(r)$. The resulting twelve-root distribution is
\begin{equation}
q(c)=\frac{\exp(\widetilde{s}_c)}{\sum_{k=0}^{11}\exp(\widetilde{s}_k)}.
\end{equation}
We compute top-1 accuracy, target rank, and $D_{\mathrm{KL}}(q^{\mathrm{clean}}\|q^{\mathrm{int}})$ from this distribution, and define the twelve-root margin as $\widetilde{s}_y-\max_{c\neq y}\widetilde{s}_c$. Thus, the twelve-root metrics compare complete root-conditioned chord continuations rather than isolated single-token logits.

\section{Data processing details}\label{app:data_processing_details}

\subsection{Nottingham}
Within each tonal segment, we select the last adjacent dominant-to-tonic root pair and trace back to the nearest preceding tonic-root chord; the excerpt spans the bars from that chord to the target, and major and minor keys use the same rule. Of 1,034 tune blocks, 54 have no valid sequence and 17 duplicate an earlier sample, leaving 963 samples. Each input is truncated immediately before the token that contains the root of the terminal tonic.

\subsection{BCMH}
\begin{figure}[H]
  \centering
  \includegraphics[width=\linewidth]{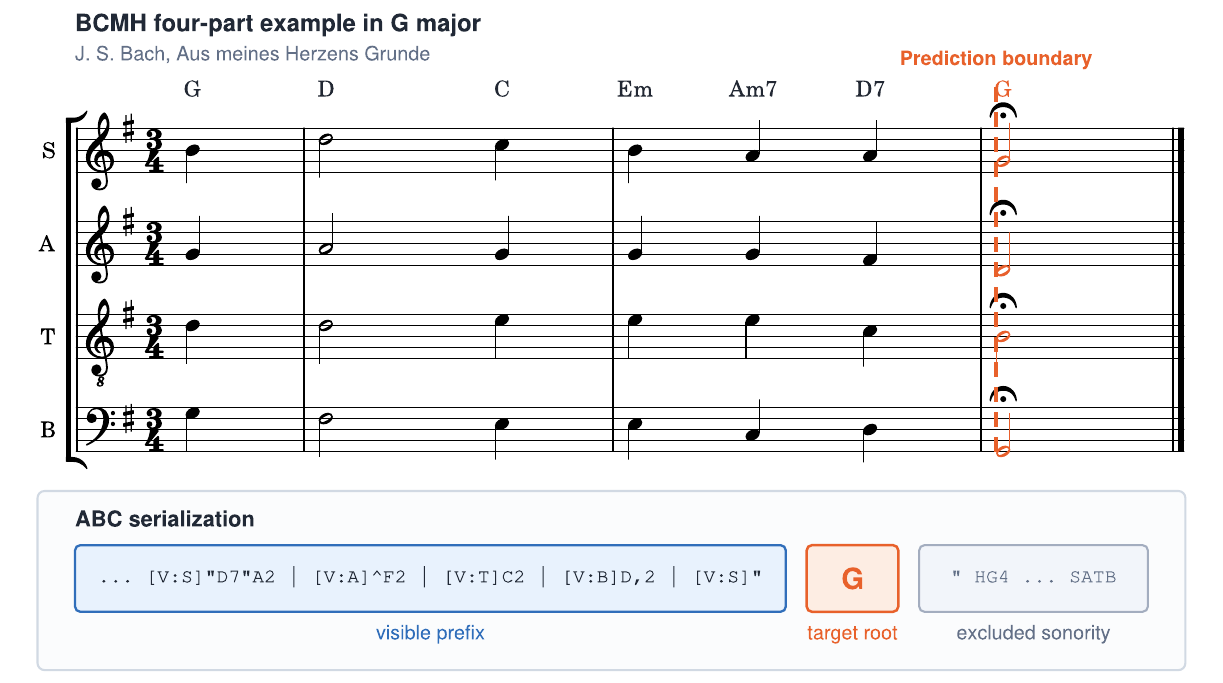}
  \caption{An SATB example from BCMH in G major, with the functional analysis $I\text{--}V^{6}\text{--}IV^{6}\text{--}vi\text{--}ii^{6}_{5}\text{--}V^{7}\text{--}I$. The orange dashed line marks the prediction boundary before the terminal G root; the preceding $V^{7}$ event is visible to the model, while the orange target sonority is excluded. The lower panel aligns this boundary with the serialized ABC span.}
  \label{fig:bcmh_case_intervention}
\end{figure}

We keep the synchronized SATB events of the 100 chorales in BCMH~\cite{BCMH_Dataset}, convert the expert Roman-numeral annotations into ABC chord symbols, and apply the same rule within each local key, keeping at most one target per chorale. Removing two exact duplicates leaves 98 samples (55 major, 43 minor). Because each chord symbol precedes its notes, the input contains all four voices up to the preceding chord but no notes of the target chord (Figure~\ref{fig:bcmh_case_intervention}).

\section{What does the circuit reveal?}\label{sec:circuit_analysis}

\subsection{Root-letter margin}\label{subsec:root_letter_margin}
We use a single-token measure. With $z(c)$ the logit of token $c$ at the prediction position and $y$ the target root letter among $\mathcal{R}=\{\texttt{A},\texttt{B},\ldots,\texttt{G}\}$, the \emph{root-letter margin} is $m=z(y)-\max_{c\in\mathcal{R}\setminus\{y\}}z(c)$; it ignores accidentals. Writing $m_s(\mathcal{A})$ for the margin of sample $s$ after ablating an edge set $\mathcal{A}\subseteq\mathcal{U}_s$, the \emph{net margin drop} compares it with the margin after ablating five same-size random edge sets $\mathcal{B}_1,\ldots,\mathcal{B}_5$:
\begin{equation}
\Delta^{\mathrm{net}}_s(\mathcal{A})=\frac{1}{5}\sum_{r=1}^{5}m_s(\mathcal{B}_r)-m_s(\mathcal{A}),
\label{eq:net_margin_drop}
\end{equation}
that is, the decrease in margin caused by ablating $\mathcal{A}$ minus the mean decrease caused by the random edge sets. Positive values mean that $\mathcal{A}$ is more damaging than random edges; we report sample means with 95\% bootstrap confidence intervals (CIs).

\subsection{The circuit is sparse}\label{subsec:sparsity}

For each sample, the \emph{edge share} is the fraction of all its cross-token edges that belong to the principal circuit, and the \emph{weight share} is the fraction of its total contribution weight that the principal circuit keeps, including the contributions of positions to themselves. The circuit contains only 1.7\% (ChatMusician) and 5.3\% (MuPT) of the cross-token edges, yet it retains about 25\% and 30\% of the contribution weight.

\subsection{The circuit is necessary for the tonic prediction across samples, modes, and models}\label{subsec:consistency}

For each sample, we measure the effect established in Section~\ref{subsec:principal_information_flow_intervention_results} by the net margin drop $\Delta^{\mathrm{net}}_s(\mathcal{E}_s)$ of its principal circuit. The \emph{consistency rate} is the fraction of samples with $\Delta^{\mathrm{net}}_s(\mathcal{E}_s)>0$. The \emph{flip rate} is the fraction of originally correct predictions, i.e., those in which the target token ranks first in the full vocabulary, that become incorrect after the ablation.

Table~\ref{tab:consistency} shows that the effect holds in almost every sample for ChatMusician and in about three quarters for MuPT.

\begin{table}[H]
  \caption{Sample-level consistency of the principal circuit intervention on Nottingham. The net margin drop $\Delta^{\mathrm{net}}_s(\mathcal{E}_s)$ is shown with its 95\% bootstrap CI; the other quantities are defined in Section~\ref{subsec:consistency}.}
  \label{tab:consistency}
  \centering
  \small
  \begin{tabular}{llrcrr}
    \toprule
    Model & Subset & $n$ & Net margin drop & Consistency rate & Flip rate \\
    \midrule
    ChatMusician-7B & All & 963 & 2.52 [2.42, 2.62] & 95.6\% & 83.2\% \\
     & Major keys & 882 & 2.43 [2.33, 2.53] & 95.6\% & 82.5\% \\
     & Minor keys & 81 & 3.49 [3.06, 3.95] & 96.3\% & 91.1\% \\
    \midrule
    MuPT-190M & All & 963 & 1.07 [0.98, 1.17] & 75.4\% & 88.3\% \\
     & Major keys & 882 & 0.99 [0.89, 1.09] & 73.9\% & 88.0\% \\
     & Minor keys & 81 & 1.97 [1.65, 2.30] & 91.4\% & 91.2\% \\
    \bottomrule
  \end{tabular}
\end{table}

\subsection{Connections inside a chord symbol matter most}\label{subsec:edge_families}

\paragraph{Edge categories.}
We call a token a \emph{chord token} if it overlaps the quoted span of a chord symbol and a \emph{non-chord token} otherwise, and we assign every edge of the principal circuit to one of five \emph{edge categories} according to its source and receiver: \emph{within one chord symbol} if both tokens lie in the same chord symbol, \emph{between chord symbols} if they lie in two different ones, \emph{chord $\rightarrow$ non-chord} or \emph{non-chord $\rightarrow$ chord} if only the source or only the receiver is a chord token, and \emph{outside chord symbols} otherwise. Edges involving a token that overlaps two adjacent chord symbols (on average 1.1 per sample in ChatMusician, none in MuPT) are excluded. The \emph{category share} of a category $f$ is $|\mathcal{E}^{f}_s|/|\mathcal{E}_s|$, where $\mathcal{E}^{f}_s\subseteq\mathcal{E}_s$ denotes its edges.

\paragraph{Equal-budget ablation.}
To compare the categories edge for edge, we ablate the same number of edges from each: $K_s=\min_f|\mathcal{E}^{f}_s|$ edges drawn uniformly from $\mathcal{E}^{f}_s$, averaged over five draws, with a random control of $K_s$ edges.

\paragraph{Results.}
Table~\ref{tab:edge_families} lists the values shown in Figure~\ref{fig:edge_family_summary}. The two models differ in what follows the chord-internal step: in ChatMusician, edges leaving a chord symbol also matter, whereas in MuPT they have small effects.

\begin{table}[H]
  \caption{Edge categories of the principal circuit on Nottingham ($n=963$). ``Share'' is the category share averaged over samples. The net margin drop is measured under the equal budget (on average 47.6 edges per category for ChatMusician and 42.5 for MuPT) and shown with its 95\% bootstrap CI.}
  \label{tab:edge_families}
  \centering
  \small
  \begin{tabular}{lrcrc}
    \toprule
    & \multicolumn{2}{c}{ChatMusician-7B} & \multicolumn{2}{c}{MuPT-190M} \\
    \cmidrule(lr){2-3}\cmidrule(lr){4-5}
    Edge category & Share & Net margin drop & Share & Net margin drop \\
    \midrule
    Within one chord symbol & 10.7\% & \phantom{$-$}1.36 [1.27, 1.44] & 4.5\% & \phantom{$-$}0.44 [0.38, 0.51] \\
    Between chord symbols & 6.7\% & \phantom{$-$}0.56 [0.48, 0.63] & 2.8\% & \phantom{$-$}0.11 [0.04, 0.18] \\
    Chord $\rightarrow$ non-chord & 8.2\% & \phantom{$-$}0.57 [0.50, 0.65] & 5.0\% & \phantom{$-$}0.05 [0.03, 0.07] \\
    Non-chord $\rightarrow$ chord & 16.3\% & \phantom{$-$}0.09 [0.05, 0.13] & 19.8\% & $-$0.13 [$-$0.16, $-$0.11] \\
    Outside chord symbols & 58.0\% & $-$0.05 [$-$0.08, $-$0.02] & 67.8\% & \phantom{$-$}0.02 [0.01, 0.03] \\
    \bottomrule
  \end{tabular}
\end{table}

\subsection{In polyphonic music, the circuit does not rely on the bass}\label{subsec:voices}

A human analyst pays particular attention to the bass, which often sounds the chord root. We group MuPT's principal edges on BCMH by the voice of their source note, $\mathcal{E}^{v}_s$ for $v\in\{\mathrm{S},\mathrm{A},\mathrm{T},\mathrm{B}\}$, together with the unions of the inner (A, T) and outer (S, B) voices and of all four voices, and ablate each set in full. The alto group matters most, the tenor group is indistinguishable from random edges, and the bass group is less damaging than random edges (Table~\ref{tab:voice_groups}). The circuit thus does not rely on the bass line in the way a human analyst would. Because BCMH has only 98 samples and the input already names each root in a chord symbol, this result remains open to further investigation.

\begin{table}[H]
  \caption{Net margin drop of the voice groups on BCMH with MuPT-190M ($n=98$), with 95\% bootstrap CIs. ``Edges'' is the mean number of ablated edges per sample. The last three rows are unions of voice groups.}
  \label{tab:voice_groups}
  \centering
  \small
  \begin{tabular}{lrc}
    \toprule
    Voice group & Edges & Net margin drop \\
    \midrule
    Soprano & 205.5 & \phantom{$-$}0.06 [0.00, 0.11] \\
    Alto & 227.4 & \phantom{$-$}0.16 [0.09, 0.22] \\
    Tenor & 214.6 & \phantom{$-$}0.01 [$-$0.05, 0.07] \\
    Bass & 229.0 & $-$0.24 [$-$0.33, $-$0.16] \\
    \midrule
    Inner voices (alto and tenor) & 442.0 & \phantom{$-$}0.16 [0.07, 0.25] \\
    Outer voices (soprano and bass) & 434.5 & $-$0.21 [$-$0.30, $-$0.12] \\
    All four voices & 876.5 & $-$0.15 [$-$0.27, $-$0.03] \\
    \bottomrule
  \end{tabular}
\end{table}

\end{document}